\documentclass[10pt,conference,a4paper]{IEEEtran}

\usepackage{todonotes}
\usepackage{enumerate}
\usepackage{geometry}
\usepackage{cite}

\ifCLASSINFOpdf
\else
\fi
\begin{document}
%
\title{Gradient Boosting Application in Forecasting of Performance Indicators Values for Measuring the Efficiency of Promotions in FMCG Retail}

\author{
	\IEEEauthorblockN{
		Joanna Henzel\IEEEauthorrefmark{1},
		Marek Sikora\IEEEauthorrefmark{6}}
	\IEEEauthorblockA{Department of Computer Networks and System\\
		Faculty of Automatic Control, Electronics and Computer Science\\
		Silesian University of Technology\\
		ul. Akademicka 16, 44-100 Gliwice, Poland\\
		Email: \IEEEauthorrefmark{1}joanna.henzel@polsl.pl,
		\IEEEauthorrefmark{6}marek.sikora@polsl.pl}
	}

\maketitle

\begin{abstract}
In the paper, a problem of forecasting promotion efficiency is raised. The authors propose a new approach, using the gradient boosting method for this task. Six performance indicators are introduced to capture the promotion effect. For each of them, within predefined groups of products, a model was trained. A description of using these models for forecasting and optimising promotion efficiency is provided. Data preparation and hyperparameters tuning processes are also described. The experiments were performed for three groups of products from a large grocery company.
\end{abstract}


%
\IEEEpeerreviewmaketitle

\section{Introduction}

Food retailing is an industry that most people have contact with. It  provides products which are necessary for everyday life. Mostly, food is bought on an ongoing basis and, because of this, precise planning of logistics, chain supplies and sales is very important. Because of the characteristics of sale of these products, they are often called  \textit{fast-moving consumer goods} (FMCG).

On the market, many retailers offering FMCG products are available, therefore it is crucial to remain competitive. One way to do this is to offer products in promotion. The importance of creating promotions in the FMCG sector can be proven by seeing the amount of money that are spent on this purpose -- in 2014 it was \$1 trillion every year as it was mentioned in \cite{Cohen2017}. Therefore, it is necessary to forecast the promotion effect and plan them with equal importance as a~regular sale.

In some cases promotions are planned based on \textit{judgmental forecasting} or using simple baseline statistical forecast with a judgmental adjustment \cite{Fildes2019b}. It means that the promotion planning process is often done manually.  However, studies have shown that using only these kinds of forecasting methods may bring bias \cite{Makridakis1986}. A better idea may be to use more advanced methods that rely mostly on knowledge that comes from historic data. Very little has been written about using Machine Learning (ML) methods for the problem of promotion optimisation and forecasting promotion effect.

The objective of this paper is to propose a new way of forecasting promotion effect using the gradient boosting method. Six different indicators are presented in order to capture the efficiency of promotions. The paper describes an advanced data preparation process. Among three groups of products, a model for each indicator was trained, examined and the optimisation of hyperparameters was conducted. The paper also describes how to use the created models in order to perform optimisation of promotions to get better outcome of the forecast.
The paper is organised as follows: the next section provides the review of literature and related works, 
section \ref{sec:problem statement} describes problem statement and presents proposed indicators.
Afterwards, the data preparation process is presented, followed by the experiments explanation.
The paper ends with some conclusions and discussion of the results.

\section{Related Works} \label{sec:related_works}

Sales forecasting plays an important part in planning and managing many commercial enterprises, including those connected with the retail sector.

Traditionally, forecasting was made using statistical methods, for example: exponential smoothing \cite{Gardner1985}, moving average and the Auto-regressive Integrated Moving Average (ARIMA) model. Well known and widely used is SARIMA -- seasonal auto-regressive integrated moving average. Some improvements of this method were proposed regarding the problem of sales forecasting in the papers \cite{Choi2011} and \cite{Arunraj2015}.

Over time, more complex methods were used and evaluated in the field of sales forecasting.  
In \cite{Chu2003} a comparison of various linear and non-linear models for this task was conducted. The best obtained model was the neural network built on deseasonalized time series data. The results suggested that non-linear models should be highly considered when dealing with modelling retail sales.
Another neural network algorithm regarding forecasting retail sales which was used for this task was back-propagation neural network (BPNN) \cite{Chen2010}. Evolutionary neural networks (ENN) were also considered in \cite{Au2008}. 
The use of the extreme learning machine (ELM) algorithm was also investigated in this area, for example in the papers \cite{Sun2008}, \cite{Xia2012} and \cite{Yu2011}. 

An important part of retail forecasting is making sales forecasts for short shelf-life food products, which are very often referred to as Fast-Moving Consumer Goods (FMCG). It is an even more complex task, because the additional products, whose sales may be overestimated, cannot be stored for a very long time in the shop. In the paper \cite{Doganis2006} a radial basis function (RBF) neural network and a designed genetic algorithm were successfully used for forecasting the sales of fresh milk. In the aspect of FMCG, the authors of \cite{Tarallo2019} showed benefits of applying Machine Learning methods in creating demand forecasting models. The use of the Autoregressive Distributed Lag model was presented in the paper \cite{Huang2014}. The authors of \cite{AdithyaGanesan2020} proposed using the Dynamic Artificial Neural Network for food sales forecasting for one of multiplexes in India.

Decision and regression tree-based methods were also taken into consideration regarding the sales forecasting. A~hybrid method of k-means algorithm and C4.5 algorithm (decision tree classifier) was shown in \cite{Thomassey2006}. In the paper \cite{Krishna2018} a~comparison of different Machine Learning Techniques was conducted regarding sales-forecasting of retail stores. The authors concluded that boosting algorithms gave better results than the regular regression ones. For them, the best results were obtained for the GradientBoost algorithm and the XGBoost implementation has been used in order to increase the accuracy.


Forecasting sales during promotions is a very challenging task as it was mentioned in \cite{Fildes2019b}. In this paper authors pointed out that usually the promotional effect was estimated by combining simple statistical forecasting methods and adding judgmental adjustment, which could lead to miscalculations.

The research about effectiveness of promotions has been conducted for a long time, mostly in the marketing research area and it is described in the practitioner literature.
This problem was raised in \cite{Blattberg1987} and \cite{Zhang2009}. The authors of \cite{Cohen2017} proposed a new formula for the promotion optimisation problem in the FMCG industry. Although these works concerned estimating the effectiveness of promotions, all of them focus on domain knowledge and do not use machine learning techniques for this task.

Multiple models for forecasting the demand during promotion periods were tested in the paper \cite{VanDonselaar2016}.
The use of PCA and pooled regression was presented in the paper \cite{Trapero2015} in  order to predict sales in the presence of promotions.
In the case of direct marketing, machine learning methods were compared and tested in the paper \cite{Cui2006}.
Interesting findings are presented in \cite{Ali2009}. The authors showed that simple statistical methods performed very well for data without promotions. For periods with promotions more advanced methods had to be used. In this paper, regression trees were used for grocery sales forecasting.

To the best of our knowledge, the tree boosting algorithm, especially the extreme gradient boosting (XGBoost) algorithm, has not been used to forecast the effect of promotions and to optimise the promotion itself. XGBoost was introduced in \cite{Chen2016}. It is a well known fact that XGBoost is highly effective for a vast range of classification and regression problems. It was, for example, used in the following areas: medicine \cite{Torlay2017}, fault detection \cite{Zhang2018}, finances \cite{Nobre2019}, accident detection \cite{Parsa2020}, and many others.

XGBoost implementation has a wide array of hyper-parameters. In order to obtain the best results, optimisation of those parameters can be performed. The most commonly used methods are random search (RS) and Bayesian Tree Parzen Estimator (TPE). These methods were used in  \cite{Wang2019} and \cite{Nishio2018}. Hyper-parameters optimisation was done using Bayesian optimisation, random search, grid search, and manual search in the paper \cite{Xia2017}.

\section{Problem statement} \label{sec:problem statement}


In different industries, promotions may have various characteristics. For example, in fashion retail it is noticeable that promotions take place mostly in specific periods during the year -- at the end of the fashion seasons.
The situation is different in grocery retail business. Multiple promotions can be observed at the same time and they are changing very rapidly. Also, alongside the regular promotions, we can distinguish promotions related to holidays and special days (e.g. Christmas, Easter or St. Valentine's Day) and discounts that are caused by upcoming expiration date.

The purpose of the promotions may be not so obvious. They should give a company bigger profit, but it is not equivalent to the willingness to sell as much as possible of a~promoted product. Of course, selling is one of the components of a~successful promotion but not the only one. For example, a~grocery retail company that set up a~promotion does not want customers to buy only the promoted product but wants clients to buy also multiple different products alongside that may be in their regular prices.

In order to capture the effectiveness of each promotion, six different indicators are proposed:

\begin{itemize}
	\item \textsc{Average number of sold units or kilograms each day} 
	(shortcut: \textsc{Avg. Amount}) -- 
	This indicator shows how many units or kilograms of the promoted product, on average, were sold during the promotion each day.
	
	\item \textsc{Average number of receipts with the promoted product} 
	(shortcut: \textsc{Avg. Nb. Receipts}) -- 
	The indicator explains in how many baskets the promoted product appeared, on average, each day during the promotion. It can be treated as an indicator of how many customers bought the product each day.
	
	\item  \textsc{Average value of a basket containing the promoted product} 
	(shortcut: \textsc{Avg. Basket}) -- 
	This indicator says what an average value of a basket was where the promoted product appeared. Assuming that customers went for shopping with the will to buy the specific product in promotion, the indicator says how much money they spent in total. The higher the indicator, the more products were bought or the more expensive products were chosen.
	
	\item \textsc{Average value of a basket containing the promoted product but disregarding the value of the promoted product}
	 (shortcut: \textsc{Avg. Basket Without Item}) -- 
	 This indicator is very similar to the previous one. It shows what an average value of a basket was where the promoted product appeared but the value of the promoted product was not taken into account. It means that this indicator is equal to 0 if the customer buys only the promoted product. 
	
	\item \textsc{Average number of unique products in the basket}
	(shortcut: \textsc{Avg. Nb. Unique Items}) -- 
	It says how varied the basket is. The higher the value of the indicator, the better -- it means that the customer not only bought a specific product but also many others.
	
	\item \textsc{Average number of the baskets}
	 (shortcut: \textsc{Avg. Nb. Clients}) -- 
	 The indicator shows how many, on average, transactions were performed each day during the promotion. It does not matter if the customer bought a~promoted product or not. 
	
\end{itemize}

The values of indicators are calculated per promotion. It means that each promotion can be described by the 6 proposed indicators.

These indicators may seem very similar, because the differences between them are very subtle. In order to show their utility, some examples are introduced:

\begin{enumerate}
	\item 100 kg of apples were sold during the promotion. The indicator \textsc{Average number of sold units or kilograms each day} tells us about it, but it does not give an information if this amount was bought by one person or by 50 people who bought 2 kg on average. This information will be provided by the \textsc{Average number of receipts with the promoted product}.
	
	\item The average value of the basket, with a product that was in promotion, was 50\$. It is the value of the indicator \textsc{Average value of a basket containing promoted product}. Now we may want to know if the rest of the products were a big part of the basket (e.g. 80 \%) or only an addition to the promoted product (e.g. 10 \% of the total value). The \textsc{Average value of a basket containing the promoted product but disregarding the value of the promoted product} gives this information. We also might want to know if the customers, on average, bought 2 unique products, that gave the value of 50 \$, or they bought 25 unique products -- the indicator \textsc{Average number of unique products in the basket} is proposed in order to capture this.
\end{enumerate}

Each of the proposed indicators are \textit{gain measures}. It means that the higher the value, the better is the promotion. They can be inversely correlated -- for example, if the price is very low, clients may buy a lot of the specific product but the diversity of products inside the basket may be very poor.

The proposed indicators describe each promotion very precisely. Knowing the value of each of them, the evaluation of the promotions can be performed. What is even more interesting, is the evaluation of future promotions so it is connected with the promotions planning. By setting up the features of the future promotion, it is possible to determine whether the predicted effect will be satisfying.

The forecasting of the promotion effect can be done for every product separately. Having the history of the promotions and their effects, we can model the characteristics of the promotion for the specific product and it is possible to predict what the effect in the future will be. Unfortunately, a number of past promotions for many products is small, so there are not many examples for training a model. 
Additionally, a question has been raised how to predict the promotion effect for a new product or an item that has never been in promotion.
One solution may be to find similar products that have similar characteristic of sales. The problem is that it is difficult to assure that this will translate to similar characteristics of promotion effect.
Another idea would be to create, based on domain knowledge, groups of products that act the same during the promotions. Then a model would be built for each of these groups. This issue, however, is out of scope of our paper.

The problem of forecasting indicators for unknown and rarely promoted products was solved by the authors -- the products were grouped by the predefined categories, e.g. vegetables, fruits, dairy products or meat. It is assumed that the products within the group will act similarly during the promotion because they are akin to each other. Therefore, it is expected that the characteristics of the indicators describing the promotion effect will be similar for products within the group.

To summarize: a new approach to the problem of forecasting the promotion effect is to calculate a model for each of the 6 proposed indicators for each predefined category (group) of products.

\section{Data preparation}

In developing models for promotions indicators and in experiments, data from a large grocery retail company were used (more than 500 stores). 
The data from groups: vegetables, fruits and dairy products were taken into account. Only regular promotions were investigated, therefore the promotions that happened before or during holidays were not included. Additionally, promotions that applied only when:
\begin{itemize}
	\item multiple units were bought (type ``buy 2 pay for 1"),
	\item minimum weight condition was met (type ``buy minimum 5 kg and get 15 \% off"),
	\item when combination of products was bought
\end{itemize}
were not taken into consideration. The same goes for products that had reduced prices because of the approaching best-before date. Also, in the examined data there were no promotions longer than 7 days. Promotions from the years 2015 to 2018 were used. Data for 2015 and part of 2016 were not completed, so there was a visibly smaller number of promotions at that period. 

One record of data described one promotion in one store. Therefore, for example, if there would be a promotion on pears in the store with ID 10 from 2018-01-22 to 2018-01-25, the record, before preparation, would look like in table \ref{table:example_record}.

\begin{table}[]
	\caption{Example of record describing promotion before preparation}
	\label{table:example_record}
\begin{tabular}{|p{0,5cm}|l|l|l|p{1,2cm}|p{1,1cm}|}
	\hline
	store \newline ID & product & start date & end date   & conditional \newline attributes & value of \newline indicator \\ \hline
	10       & pears     & 2018-01-22 & 2018-01-25 & ...            & 123.56             \\ \hline
\end{tabular}
\end{table}

\subsection{Attributes} \label{subsec:attributes}

In the research, extended numbers of conditional attributes were taken into consideration when preparing data sets. A few main categories of the attributes can be distinguished:

\begin{itemize}
	\item connected with price,
	\item connected with the time and duration of the promotion,
	\item describing the advertisement media (promotion channels),
	\item describing the store and its surroundings,
	\item describing the impact of other promotions.
\end{itemize}

In the first category, only 2 attributes were included: the price of a product and a change of the price.

Time attributes connected with the promotion were:

\begin{itemize}
	\item number of days of the promotion,
	\item weekday of the first day of the promotion,
	\item attributes created based on the date of the first day of the promotion: year number, month number, day number, week number, number of a day in the year, and the season.
\end{itemize}

Considering information about promotion channels, binary attributes were added. They described if the promotion was advertised on  TV, on the radio, on the Internet or in a different way.

Additionally, new variables describing combinations of the promotion channels were added to the data sets. For each combination, new attributes were created as a result of binary operations AND, OR and XOR (only when combination consisted of 2 elements). For example, if the undermentioned statements, were true, then a new variable got value 1, otherwise -- 0.

\begin{itemize}
	\item Promotion was on TV or on the radio. (OR operation)
	\item Promotion was on TV or on the radio or on the Internet. (OR operation)
	\item Promotion was on the TV and on the radio. (AND operation)
	\item Promotion was either on the Internet or on the radio. (XOR operation)
\end{itemize}

We can assume that promotions in similar stores (for example in small villages or in big cities) can have similar characteristics. For example, the customers in a rich city buy more expensive products in general, therefore the value of the basket is automatically higher than in other stores. 
The exemplary attributes that were used in order to capture these characteristics were:

\begin{itemize}
	\item number of inhabitants within 1 km,
	\item number of inhabitants per 1 square km,
	\item number of inhabitants within a 5-minute driving range,
	\item unemployment rate,
	\item number of cars per 1,000 inhabitants,
	\item average monthly salary,
	\item tourism ratio, etc.
\end{itemize}

The last but not least, attributes connected with the impact of other promotions were added. As it was mentioned in the section \ref{sec:problem statement}, promotions rarely ever take place one at a time. It is a possible situation, that a client that bought the considered product came to the store because of another promotion. It is impossible to capture clients' intentions fully, but it can be assumed that the more promotions in the shop, the more clients will come. Because of this, the following attributes were added to the data:

\begin{itemize}
	\item Number of all promotions in a store.
	\item Number of all promotions that were advertised on TV, radio or internet.
	\item Number of all promotions that were advertised on TV, radio, internet or in a different way.
\end{itemize}


\subsection{Matching periods without promotions}

In order to capture the characteristics of products in the group, matching records without promotions were found for most of the records in the data set. The matching period had to meet the following conditions:

\begin{itemize}
	\item It considered the same product as the promotion.
	\item It considered the same store.
	\item It had to last as many days as the considered promotion.
	\item It had to start on the same weekday as the promotion.
	\item The considered product was not in promotion on any given day.
	\item The period without promotion could occur maximum 4~weeks and minimum 1 week before the promotion.
\end{itemize}

The matching period was not found for all promotions because of the lack of meeting the requirements.

\begin{figure}[!t]
\centering
\includegraphics[width=3.0in]{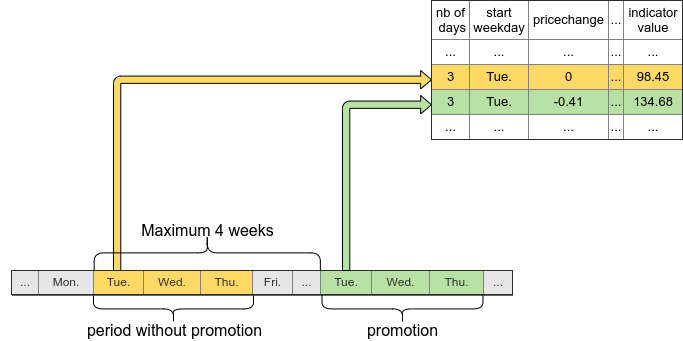}
\caption{Finding matching record without promotion}
\label{fig:matching_periods}
\end{figure}

The illustration of finding the matching periods was shown in figure \ref{fig:matching_periods}.

In the final data sets, records connected with periods without promotions were distinguished from promotions by having 0~value in an attribute describing the change of a price.

\subsection{Standardisation of the indicators} \label{subsec:standardisation}

The standardisation of two proposed indicators was performed. These were:

\begin{itemize}
	\item \textsc{Average number of sold units or kilograms each day} and
	\item \textsc{Average number of receipts with the promoted product}.
\end{itemize}

The z-score standardisation was used, but for each product and each store separately. The reason for using standardisation for those indicators was that they were referring to the specific values connected with the sale characteristics of a considered product. For example, it is predictable that during promotions with 20\% reduction, apples will be sold more than pomelos, because apples are cheaper and they are bought more often in general. The values of the indicator \textsc{Average number of sold units or kilograms each day} will be from a different range for those products. 
This does not mean, however, that the impact of the 20\% reduction does not affect in the same way the increase of sold units of apples and pomelos. In order to capture the general characteristics of products in a group, the standardisation of those indicators was performed.

\section{Experiments}

The experiments of the proposed solution for problem of forecasting the promotion effect were conducted for the following categories of products: fruits, vegetables and dairy products.
For each category and each proposed indicator, a forecasting model was constructed.
In training data sets, records from 2015-2017 describing promotions and matching periods without promotions were included. In test data sets, records with promotions from 2018 were used.
For all indicators within one group of products, conditional attributes in data were the same (described in subsection \ref{subsec:attributes}). The decision attributes were the values of the considered indicators. 

When testing models, cross-validation was not performed. The reason for this is the fact that although the data sets were not typical time-series data, the records could be set in chronological order. Using cross-validation, the testing of a~model might be performed on records preceding the training data. 


XGBoost (eXtreme Gradient Boosting)~\cite{Chen2016} from the R~package \texttt{xgboost} \cite{rpackage:xgboost} implementation was used for training forecasting models. This gradient boosting framework was chosen because it is a well-known method, which get very good results when working with table-structured data. For example, among the 29 challenges winning solutions posted on a machine learning competition site named Kaggle in 2015, 17 solutions used XGBoost \cite{Chen2016}.
The experiments described in this paper were also based on tabular data, therefore using XGBoost was a justified idea.
Additionally, the paper  \cite{Krishna2018} showed that this algorithm has given the best results for sales-forecasting of retail stores in their experiments, so it was very likely to give good results also for the problem of forecasting the promotion effect in retail sector.
In order to evaluate the models efficiency, the following error measures were used:

\begin{itemize}
	\item \textit{Mean Absolute Error (MAE)}: 
	\[ MAE = \frac{\sum_{i=1}^{n}|F_i - A_i|}{n} \]
	
	\item \textit{Root Mean Square Error (RMAE)}: 
	\[ RMSE = \sqrt{\frac{\sum_{i=1}^{n}(F_i - A_i)^2}{n}} \]
	
	\item \textit{Mean Absolute Percentage Error (MAPE)}: 
	\[ MAPE = \frac{1}{n}\sum_{i=1}^{n} \left| \frac{A_i - F_i}{A_i}\right| \]
	
	\item \textit{Weighted Mean Absolute Percentage Error (WMAPE)}: \\
	\[ WMAPE = \frac{\sum_{i=1}^{n}\mid A_i - F_i\mid}{\sum_{i=1}^{n} A_i} \]
\end{itemize}
where $A_i$ is the actual value and $F_i$ is the forecast value.

\begin{table}[]
	\caption{Results of models effectiveness using default hyperparameters}	
	\label{table:default_results}
	\centering
	\scalebox{0.75}{
		\begin{tabular}{|l|l|l|l|l|l|}
			\hline
			category & indicator & MAE & RMSE & MAPE & WMAPE \\ 
			\hline
			dairy products & AVG. AMOUNT & 12.35 & 19.49 & 0.51 & 0.38 \\ 
			dairy products & AVG. NB. RECEIPTS & 5.75 & 8.15 & 0.44 & 0.33 \\ 
			dairy products & AVG. BASKET & 14.95 & 21.53 & 0.19 & 0.18 \\ 
			dairy products & AVG. BASKET WITHOUT ITEM & 14.41 & 20.97 & 0.18 & 0.19 \\ 
			dairy products & AVG. NB. UNIQUE ITEMS & 2.12 & 2.86 & 0.14 & 0.14 \\ 
			dairy products & AVG. NB. CLIENTS & 165.67 & 247.54 & 0.10 & 0.10 \\ 
			fruits & AVG. AMOUNT & 44.57 & 84.33 & 1.18 & 0.51 \\ 
			fruits & AVG. NB. RECEIPTS & 27.92 & 45.22 & 0.87 & 0.39 \\ 
			fruits & AVG. BASKET & 18.79 & 26.64 & 0.19 & 0.20 \\ 
			fruits & AVG. BASKET WITHOUT ITEM & 17.40 & 25.09 & 0.19 & 0.20 \\ 
			fruits & AVG. NB. UNIQUE ITEMS & 2.26 & 3.17 & 0.13 & 0.14 \\ 
			fruits & AVG. NB. CLIENTS & 135.50 & 178.62 & 0.08 & 0.08 \\ 
			vegetables & AVG. AMOUNT & 24.37 & 44.89 & 0.48 & 0.35 \\ 
			vegetables & AVG. NB. RECEIPTS & 21.04 & 37.49 & 0.42 & 0.33 \\ 
			vegetables & AVG. BASKET & 18.31 & 26.29 & 0.18 & 0.19 \\ 
			vegetables & AVG. BASKET WITHOUT ITEM & 18.10 & 25.53 & 0.19 & 0.20 \\ 
			vegetables & AVG. NB. UNIQUE ITEMS & 2.34 & 3.24 & 0.13 & 0.14 \\ 
			vegetables & AVG. NB. CLIENTS & 171.61 & 229.00 & 0.10 & 0.10 \\ 
			\hline
		\end{tabular}
	}
\end{table}

\begin{table}[]
	\caption{Results of models effectiveness after hyperparameters optimisation}	
	\label{table:after_optimalization}
	\centering
	\scalebox{0.75}{
		\begin{tabular}{|l|l|l|l|l|l|}
			\hline
			category & indicator & MAE & RMSE & MAPE & WMAPE \\ 
			\hline
			dairy products & AVG. AMOUNT & 12.31 & 18.71 & 0.53 & 0.38 \\ 
			dairy products & AVG. NB. RECEIPTS & 5.75 & 8.08 & 0.45 & 0.33 \\ 
			dairy products & AVG. BASKET & 13.93 & 20.14 & 0.17 & 0.17 \\ 
			dairy products & AVG. BASKET WITHOUT ITEM & 14.26 & 20.32 & 0.19 & 0.18 \\ 
			dairy products & AVG. NB. UNIQUE ITEMS & 2.04 & 2.72 & 0.14 & 0.13 \\ 
			dairy products & AVG. NB. CLIENTS & 129.75 & 177.15 & 0.08 & 0.08 \\ 
			fruits & AVG. AMOUNT & 39.72 & 74.62 & 1.11 & 0.45 \\ 
			fruits & AVG. NB. RECEIPTS & 24.87 & 39.39 & 0.85 & 0.35 \\ 
			fruits & AVG. BASKET & 15.29 & 22.44 & 0.16 & 0.16 \\ 
			fruits & AVG. BASKET WITHOUT ITEM & 14.73 & 21.78 & 0.17 & 0.17 \\ 
			fruits & AVG. NB. UNIQUE ITEMS & 1.84 & 2.60 & 0.12 & 0.11 \\ 
			fruits & AVG. NB. CLIENTS & 125.15 & 164.49 & 0.07 & 0.07 \\ 
			vegetables & AVG. AMOUNT & 22.97 & 42.42 & 0.47 & 0.33 \\ 
			vegetables & AVG. NB. RECEIPTS & 19.52 & 34.78 & 0.41 & 0.31 \\ 
			vegetables & AVG. BASKET & 14.39 & 21.56 & 0.15 & 0.15 \\ 
			vegetables & AVG. BASKET WITHOUT ITEM & 14.63 & 21.65 & 0.16 & 0.16 \\ 
			vegetables & AVG. NB. UNIQUE ITEMS & 1.89 & 2.71 & 0.11 & 0.11 \\ 
			vegetables & AVG. NB. CLIENTS & 135.57 & 178.51 & 0.08 & 0.08 \\ 
			\hline
		\end{tabular}
	}
\end{table}

The mean absolute percentage error (MAPE) is very intuitive and easy to interpreted, however it is meaningful only when the values are large. If the actual value is close to 0, the value of MAPE is approaching infinity and it gives uninterpreted results. In order to bypass these disadvantages, a similar measure -- WMAPE -- was used. It is the sum of absolute errors divided by the sum of the actual values and it works well with smaller numbers. It is widely used in the retail sector.

Firstly, the XGBoost method was used with default hyperparameters. The results, obtained for test data sets, are presented in table \ref{table:default_results}.
For two indicators that were standardised (see subsection \ref{subsec:standardisation}), error measures were calculated after changing forecasted, standardised values to the real values.

\subsection{Optimisation}

The optimisation of hyperparameters was performed for each created model. A \textit{grid search} method was used.
Six hyperparameters were optimised:

\begin{itemize}
	\item \textit{nrounds} -- maximum number of boosting iterations; range:~$[1, \infty )$.
	\item \textit{base\_score} -- the initial prediction score of all instances; range:~$(-\infty, \infty )$.
	\item \textit{eta} -- boosting learning rate; range: $[0, 1]$.
	\item \textit{gamma} -- minimum loss reduction required to make a~further partition on a leaf node of the tree; range: $[0, \infty)$.
	\item  \textit{max\_depth} -- maximum depth of a tree; range: $[1, \infty)$.
	\item \textit{subsample} -- subsample ratio of the training instance; range: $(0, 1]$.
\end{itemize}
A detailed description of the above parameters can be found in \cite{rpackage:xgboost}.

\begin{figure}[]
	\centering
	\includegraphics[height=22.5cm]{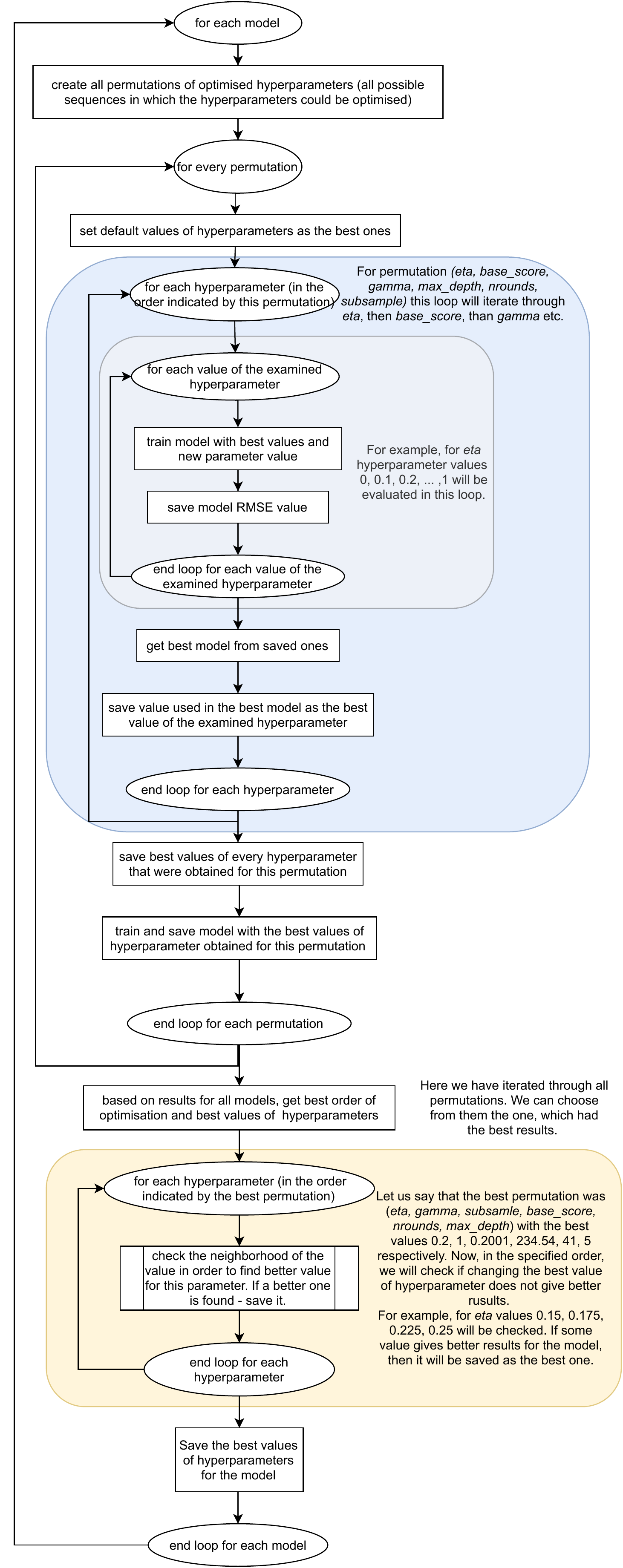}
	\caption{Flowchart of the hyperparameter optimisation process.}
	\label{fig:flowchart}
\end{figure}

In the beginning, all possible sequences in which the hyperparameters could be optimised were determined. Six parameters were used, so 720 permutations were obtained.  For example, the first permutation was \textit{eta, base\_score, gamma, max\_depth, nrounds, subsample} -- it means that at first the \textit{eta} hyperparameter was optimised, then \textit{base\_score}, afterwards \textit{gamma} and so on.
In each permutation, each hyperparameter was changed several times in order to find the best value. The table \ref{table:hyperparameters} shows values that were used in this process.
After iterating through each hyperparameter, the best set of the hyperparameters values of the specific permutation was obtained.
Having results for 720 permutations, the best among them was chosen.
After this step, the best order of optimising the parameters and the best values for them were determined.
In the end, the neighbourhood of the examined hyperparameters values were searched. It was performed in the order determined in the previous step (the order of the best permutation).
The optimisation was performed using the validation set that was extracted from the training data set. 
The flowchart of the described optimisation process is shown in figure \ref{fig:flowchart}. The RMSE measure was used as the optimisation criterion.

\begin{table}[]
	\caption{Values of hyperparameters used in optimisation process}	
	\label{table:hyperparameters}
	\centering
	\begin{tabular}{|l|p{6cm}|}
		\hline
		\multicolumn{1}{|c|}{hyperparameter} & tested values                                                                                                                                   \\ \hline
		\textit{nrounds}                      & 1,  21,  41,  61,  81, 101, 121, 141, 161, 181, 201                                                                                             \\ \hline
		\textit{base\_score}                  & Depending on indicator values. Calculated as 11 quantiles from indicator values with the following probabilities: 0.0, 0.1, 0.2, ..., 0.9, 1.0. \\ \hline
		\textit{eta}                          & 0.0,  0.1,  0.2,  0.3,  0.4,  0.5,  0.6,  0.7,  0.8,  0.9,  1.0                                                                                 \\ \hline
		\textit{gamma}                        & 0, 1, 2, 3, 4, 5, 6, 7, 8, 9, 10                                                                                                                \\ \hline
		\textit{max\_depth}                   & 1, 4, 7, 10, 13                                                                                                                                 \\ \hline
		\textit{subsample}                    & 0.0001, 0.1001, 0.2001, ..., 0.9001                                                                                                             \\ \hline
	\end{tabular}
\end{table}

The results of models efficiency, calculated for the test data sets after hyperparameters optimisation, were shown in table \ref{table:after_optimalization}. 
It can be observed that for most of the models metrics, the optimisation has given better results than for default models. The details can be seen by comparing table \ref{table:default_results} and table \ref{table:after_optimalization}. 
The optimisation was carried out based on the RMSE measure. The improvement of this metric was observed for every examined case. The table \ref{table:RMSE_difference} shows the exact results.

\begin{table}[]
	\caption{RMSE improvement after hyperparameter optimisation. $RMSE_{diff}$ is a difference of RMSE before optimisation (table~\ref{table:default_results}) and RMSE after optimisation (table~\ref{table:after_optimalization}).}	
	\label{table:RMSE_difference}
	\centering
	\scalebox{0.93}{
	\begin{tabular}{|l|l|l|l|}
		\cline{2-4}
		\multicolumn{1}{l|}{} & \multicolumn{3}{c|}{$RMSE_{diff}$} \\ \hline
		indicator & dairy products & fruits & vegetables \\ 
		\hline
		AVG. AMOUNT & 0.78 & 9.72 & 2.47 \\ 
		AVG. NB. RECEIPTS & 0.07 & 5.83 & 2.71 \\ 
		AVG. BASKET & 1.38 & 4.19 & 4.74 \\ 
		AVG. BASKET WITHOUT ITEM & 0.65 & 3.31 & 3.88 \\ 
		AVG. NB. UNIQUE ITEMS & 0.13 & 0.56 & 0.53 \\ 
		AVG. NB. CLIENTS & 70.39 & 14.13 & 50.49 \\ 
		\hline
	\end{tabular}
	}
\end{table}

\section{Conclusion and Discussion}

Promotions play an important role in the retail sector. When performed suitably, they can give a company bigger profit and bring in more clients to the store. 


This study has attempted to introduce a new method of planning and forecasting future promotions using the XGBoost algorithm. Six unique indicators that measure the promotion efficiency were proposed in this paper. These indicators not only describe the sale of a specific product, but characterise the promotions in a much more profound way. Being able to forecast the value of each of them, promotions can be better planned. Indicators forecasts give  information if the future promotion, with the given characteristics, like change of price or the weekday when it should start, is likely to be performed satisfactorily. If not, better attributes can be chosen. 

In the paper the authors described the data sets preparation process with the use of extended and precisely chosen attributes that could be not so obvious to use.
The authors also proposed a solution for forecasting the promotion effect for new, unknown products or products with a small number of past promotions. The models were developed for groups of products and not for each product separately. The experiments were performed for 3 groups: vegetables, dairy products and fruits.
A model using XGBoost was developed for each indicator and each group of products. Additionally, the hypermarameters optimisation was performed in order to obtain better models accuracy. 
It is worth emphasizing that such optimisation can be carried out for any error measure.

\begin{figure}[!t]
	\centering
	\includegraphics[width=3.3in]{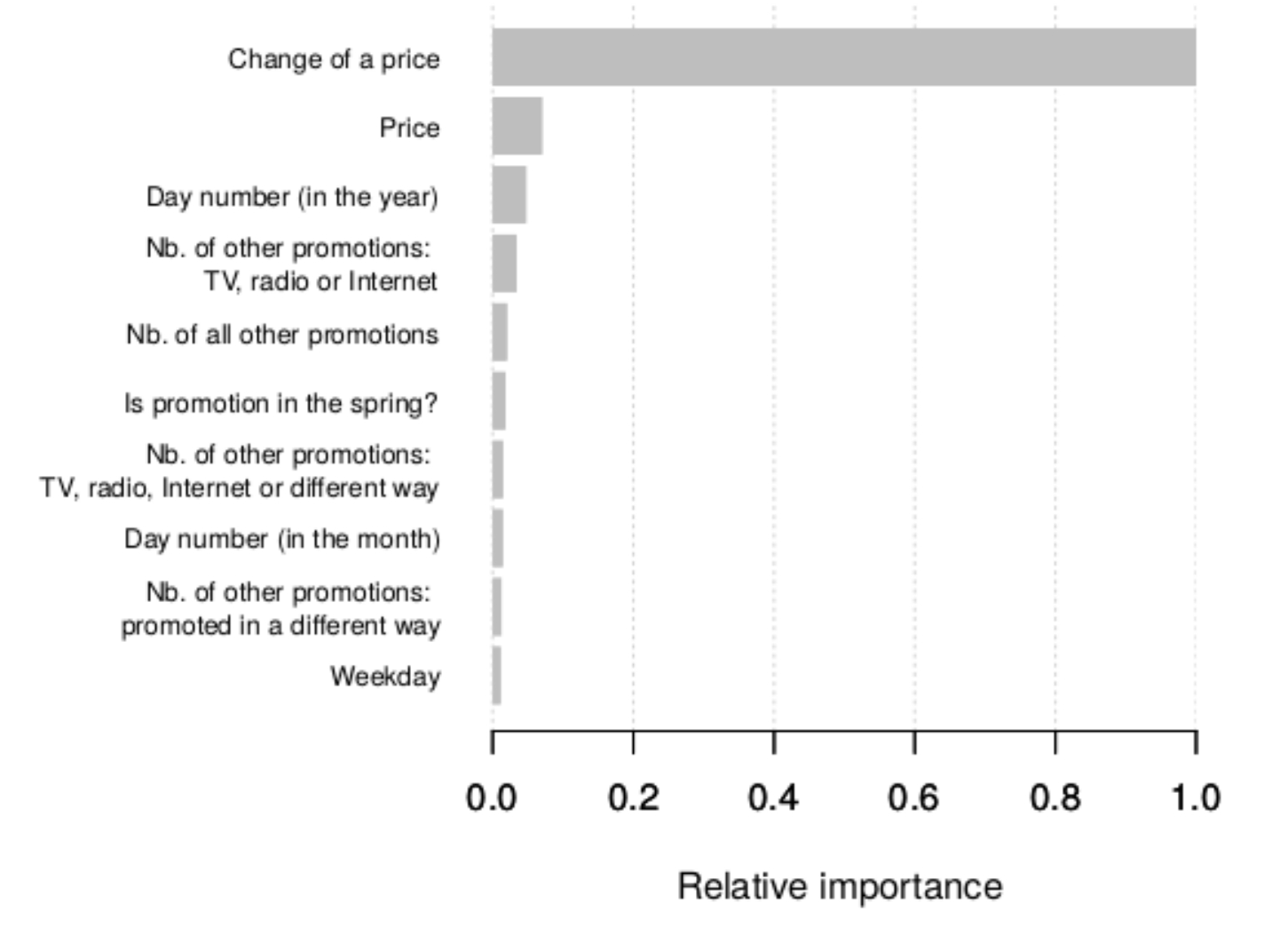}
	\caption{Plot of feature importance for the model of the indicator AVG.~AMOUNT for dairy products.  most important features are shown and the important values are represented as relative to the highest ranked feature.}
	\label{fig:31_avg_amount_importance}
\end{figure}

The created models provide also a description of the features importance. Figure \ref{fig:31_avg_amount_importance} shows a plot of 10 most important attributes of the model trained for indicator \textsc{Avg. Amount} and dairy products. It can be observed that the change of a price and the price itself are the most important features that influence the amount of sold units during the promotions for this model. In the process of planning promotions, when the results of forecast are not satisfactory, one can tune, starting from these 2 attributes, the promotions characteristics in order to get better results. After making changes in the planned promotions, the predictions can be performed again. If the results are still not satisfying, the previous steps can be repeated. This way the process of optimising future promotions can be performed.

Five most important features for each indicator are presented below. The order, in which the attributes are listed below, was obtained by calculating average importance score of each feature taking into account the results of each group of products:

\begin{itemize}
	\item \textsc{Avg. Amount}:
	change of a price; 
	day number (in the year); 
	price; 
	number of all promotions that are happening in the store and are advertised on TV, radio or Internet; 
	day number (in the month).

	\item \textsc{Avg. Nb. Receipts}:
		change of a price; 
		number of competitors; 
		number of inhabitants within a 10-minute driving range; 
		number of inhabitants within 1 km; 
		number of inhabitants within 500 m.
		
	\item \textsc{Avg. Basket}:
		price;
		number of inhabitants within 500 m;
		change of a price;
		day number (in the year);
		 weekday.
		 
	\item \textsc{Avg. Basket Without Item}:
		price;
		number of inhabitants within 500 m;
		change of a price;
		day number (in the year);
		 weekday.

	\item \textsc{Avg. Nb. Unique Items}:
		number of inhabitants within 500 m;
		price;
		change of a price;
		weekday;
		distance from a competitor.
		
	\item \textsc{Avg. Nb. Clients}:
		number of inhabitants within 500 m;
		number of inhabitants within 1 km;
		number of inhabitants within a 5-minute driving range;
		purchasing rate;
		tourism ratio.
\end{itemize}

As it can be observed, not all features are possible to change in the process of the promotions planning. However, the ranking may suggest the order in which attribute values should be tuned to get better forecasting results. The most important features for \textsc{Avg. Nb. Clients} are not connected with promotions, so the conclusion can be drawn that this indicator is little affected by them.

Summarising the practical aspect of the research: using the presented methodology it is possible to train models for forecasting promotion efficiency. At the input of the models, the features of the future promotion are placed, including change of a price, promotion channels, store attributes and a~number of days of the promotion. At the output of the models, the values of the indicators are obtained. They give information on whether the promotion will be successful.

The challenge for future research will be to investigate the efficiency of multi-target prediction methods for the problem of forecasting all six proposed indicators.

In conclusion, this paper has shown a new way of planning and forecasting promotions using Machine Learning techniques. This, to our knowledge, is the first study to examine the utility of the Gradient Boosting method in the problem of forecasting the future promotion effect.


\section*{Acknowledgment}


This work was partially supported by the European Union through the European Social Fund (grant POWR.03.05.00-00-Z305).
The work was carried out in part within the project co-financed by European Funds entitled ``Decision Support and Knowledge Management System for the Retail Trade Industry (SensAI)" (POIR.01.01.01-00-0871/17-00).



%
%
%

\bibliographystyle{IEEEtran}
\bibliography{bibliography/retail_forecasting,bibliography/xgboost,bibliography/promotions_forecasting}

\end{document}